\documentclass[preprint]{aastex}
\begin{document}
\newcommand{\calu}{{\cal U}}
\newcommand{\calq}{{\cal Q}}
\newcommand{\bx}{{\rm \bf x}}
\newcommand{\bk}{{\bar{\kappa}}}
\title{Detection of dark matter Skewness in the VIRMOS-DESCART survey:
Implications for $\Omega_0$} 
\author{Ue-Li Pen}
\affil{Canadian Institute for Theoretical Astrophysics, University of
Toronto, M5S 3H8, Canada; pen@cita.utoronto.ca}
\author{Tongjie Zhang}
\affil{Department of Astronomy, Beijing Normal University, Beijing 100875, 
P.R.China;
tjzhang@bnu.edu.cn; and 
Canadian Institute for Theoretical Astrophysics, University of Toronto,M5S 
3H8,Canada;
tzhang@cita.utoronto.ca}
\author{Ludovic van Waerbeke}
\affil{Institut d'Astrophysique de Paris, 98 bis, boulevard Arago, 75014
Paris, France; waerbeke@iap.fr}
\author{Yannick Mellier}
\affil{Institut d'Astrophysique de Paris, 98 bis, boulevard Arago, 75014
Paris, France; mellier@iap.fr}
\affil{Observatoire de Paris/LERMA, 77 avenue Denfert Rochereau, 75014
Paris, France}
\author{Pengjie Zhang}
\affil{Department of Astronomy and Astrophysics, University of
Toronto, M5S 3H8, Canada;  
zhangpj@cita.utoronto.ca} 
\author{John Dubinski}
\affil{Canadian Institute for Theoretical Astrophysics, University of
Toronto, M5S 3H8, Canada, dubinski@cita.utoronto.ca} 

\begin{abstract}

Weak gravitational lensing provides a direct statistical measure of
the dark matter distribution.  The variance is easiest to measure,
which constrains the degenerate product $\sigma_8 \Omega_0^{0.6}$.
The degeneracy is broken by measuring the skewness arising from
the fact that densities must remain positive, which is not possible when
the initially symmetric perturbations become non-linear.  Skewness measures
the non-linear mass scale, which in combination with the variance measures
$\Omega_0$ directly.  We present the first detection of dark matter skewness
from the Virmos-Decart survey.  We have measured the full three point
function, and its projections onto windowed skewness.  We separate the
lensing mode and the $B$ mode.  The lensing skewness is detected for a
compensated Gaussian on scales of 5.37 arc minutes to be $\bk^3 = 1.06\pm
0.06\times 10^{-6}$.  The B-modes are consistent with zero at this scale.
The variance for the same window function is  $ \bk^2= 5.32\pm 0.62\pm
0.98 \times 10^{-5}$, resulting in $S_3=375^{+342}_{-124}$.  Comparing to
N-body simulations, we find $\Omega_0<0.5$ at 90\% confidence.  The
Canada-France-Hawaii-Telescope legacy survey and newer simulations
should be able to improve significantly on the constraint.
\end{abstract}


\section{Introduction}

A direct measurement of the mass distribution in the universe has been a
major challenge and focus of modern cosmology.  The deflection of light
by gravity results in the gravitational lensing effect, which is a direct
measure of the strength of gravitational clustering.  Recently several
groups have been able to measure this weak gravitational lensing
effect
\citep{2002astro.ph..3134B,2002ApJ...572L.131R,2002ApJ...572...55H,2002A&A...393..369V,2002astro.ph.10604J,2002astro.ph.10213B,2002astro.ph.10450H}.
Due to the weakness of the effect, all detections have been statistical in
nature, primarily in regimes where the signal-to-noise is less than unity.

One of the degeneracies in the measurement of the power spectrum is
that between the normalization of the amplitude of fractional
fluctuations parameterized by $\sigma_8$, the linearly extrapolated
standard deviation in spheres of radius $8 h^{-1}$ Mpc, and the
present day density of matter $\Omega_0$.  One measures the
fluctuations in the projected mass density $\kappa$, which can either
arise by a large total mass density with small fluctuations, or a
smaller mean mass density with larger fractional fluctuations.
Typically one constrains the combinations $\sigma_8
\Omega_0^{0.55}$\citep{2002A&A...393..369V}.  A very similar degeneracy
arises in all dynamical measurements of mass, including redshift space
distortions \citep{1987MNRAS.227....1K}, cluster abundance
\citep{1998ApJ...498...60P}, galaxy peculiar velocities, and pairwise
velocities \citep{1983ApJ...267..465D}.

In the standard model of cosmology, fluctuations start off small,
symmetric and Gaussian.
As fluctuations grow by gravitational instability, this symmetry can
no longer be maintained: overdensities can be arbitrarily large, while
under dense regions can never have less than zero mass.  This leads
to a skewness in the distribution of matter fluctuations.  The effect
has been studied theoretically \citep{1997A&A...322....1B} and initial
detections have been reported \citep{2002A&A...389L..28B}.  In second
order perturbation theory, one finds that the skewness scales as the
square of the variance, and inversely to density.  In terms of the
dimensionless surface density $\kappa$ and source redshift $z_s$, one has
\begin{equation}
S_3 \equiv \frac{\langle \bk^3\rangle}{\langle \bk^2\rangle^2} 
\propto \Omega_0^{-0.8} z_s^{-1.35}.
\label{eqn:s3t}
\end{equation}
$S_3$ does not depend on the power spectrum normalization, but does
require knowledge of the source distribution.

The detection of this effect in real data is challenging.  The shear
field is only sampled at highly irregularly spaced galaxy positions.
Just to accurately measure the effective windowed variance of the
density field one needs to resort to the two point correlation function
\citep{2002ApJ...567...31P} which can be done cheaply computationally, or
through an optimal estimator which requires sophisticated algorithms for
the large data sets \citep{2002astro.ph.10478P}.  For low signal-to-noise
data, the two point correlation function is the optimal procedure to
measure windowed variances and power spectra.

To measure the skewness of a distribution sampled on
an irregular pattern requires measuring the three point function.
This is itself a computationally complex tasks for the spin 2 weak
lensing shear field.  \cite{2003A&A...397..405B} explored a simplified
approach using specific geometrical configurations in the shear
pattern. They applied their method to the Virmos-Descart data and
found the amplitude and the shape of their 3-point shear correlation
function over angular scales ranging from one to five arc-minutes follow
theoretical expectation of popular cosmological CDM models. Although
their detection is strong (4.9 $\sigma$), the cosmological interpretation
of non-linear features together with the properties of the 2-point shear
correlation function is difficult. In particular, one does not know
yet how it can be used to break the $\Omega_0$-$\sigma_8$ degeneracy,
as the skewness of the convergence field can do. Hence, despite recent
theoretical investigations of the 3-point shear correlation function
\citep{2002astro.ph..7454S,2002astro.ph..8075Z,2002astro.ph.10261T},
the projection of the three point function onto the skewness of the
smoothed convergence is still the easiest and most direct way to constrain
cosmological scenario with cosmic shear data.

In this paper we present the first measurement of dark matter skewness
from the Virmos-Descart survey\footnote{http://terapix.iap.fr/Descart}.
Despite the strong inhomogeneities in the galaxy distribution produced
by the masking, we found an optimal weighting scheme that allows us to
compute a robust and reliable skewness estimator.  The skewness and
variances are computed with the same compensated Gaussian smoothing
window, and a full separation into modes has been accomplished.  We have
performed N-body simulations to calibrate the results.

\section{Data}

The Virmos-Descart data consist in four uncorrelated patches (referred
as field F02, F10, F14 and F22 according to their RA position) of
about 4 square-degrees each and separated by more than 40 degrees. The
fields have been observed with the CFH12k panoramic CCD camera,
mounted at the Canada-France-Hawaii Telescope prime focus, over the
periods between January 1999 and November 2001.

The observations and data reduction have been
described in previous Virmos-Descart cosmic shear papers
\citep{2000A&A...358...30V,2001A&A...374..757V,2002A&A...393..369V}.
A detailed discussion on the data quality will be presented
elsewhere (McCracken et al in preparation), but most of the
data properties relevant for cosmic shear are already presented in
\citet{2000A&A...358...30V,2001A&A...374..757V,2002A&A...393..369V}.
Shortly, the observations have been done with the I-band filter
available on the CFH12k camera with typical exposure time of one hour.
The median seeing of the data set is 0.85 arc-second and the limiting
depth is $I_{AB}=24.5$. All data were processed at the Terapix data
center\footnote{http://terapix.iap.fr}.  Although the data observed during
that period cover 11.5 sq-degrees, a significant fraction of the field
is lost by the masking procedure, so the final area covered by the data
discussed in this work is 8.5 sq-degrees.

Catalog construction, galaxy selection, PSF anisotropy
correction and shape measurements are presented in
\citet{2000A&A...358...30V,2001A&A...374..757V,2002A&A...393..369V}.
The final cosmic shear catalog contains 392,055 galaxies with
magnitude $I_{AB}>22$ with median $I_{AB}=$23.6.   Several careful
checks have demonstrated that systematic residuals are very
small. However, \citet{2001A&A...374..757V,2002A&A...393..369V}
and \citet{2002ApJ...567...31P} have shown that a B-mode signal
still remains on scales larger than 10 arc-minutes. Its origin
is not yet understood.  Below that angular scale, its amplitude
is smaller than the E-mode, confirming the gravitational nature
of the signal.  \citet{2001A&A...374..757V,2002A&A...393..369V}
and \citet{2002A&A...389L..28B} have used these catalogs to measure
the 2-point and 3-points shear correlation functions and constrain
cosmological models.

The method can be sumarized as the
following\citet{2001A&A...374..757V}: The HDF samples are complete up
to the $I_{AB}=27^{\rm th}$ magnitude, and can therefore be used to
calibrate the redshift distribution of our VIRMOS sample. In practice,
we identify each redshift bin $n_{m_i}(z)$ of the HDF's corresponding
to each magnitude bins $m_i$ (magnitude bins have a fixed width of
$\Delta_m=0.5$). The redshift distribution of the VIRMOS sample is
estimated by adding up the $n_{m_i}(z)$ with a proper weighting
$w_{m_i}$.  The weight $w_{m_i}$ corresponds to the ratio of number
counts, per magnitude bin, of the VIRMOS and the HDF catalogues. The
completeness of VIRMOS (which occurs 2.5 magnitudes brighter than the
HDF) is then properly taken into account.

\begin{figure}
\plotone{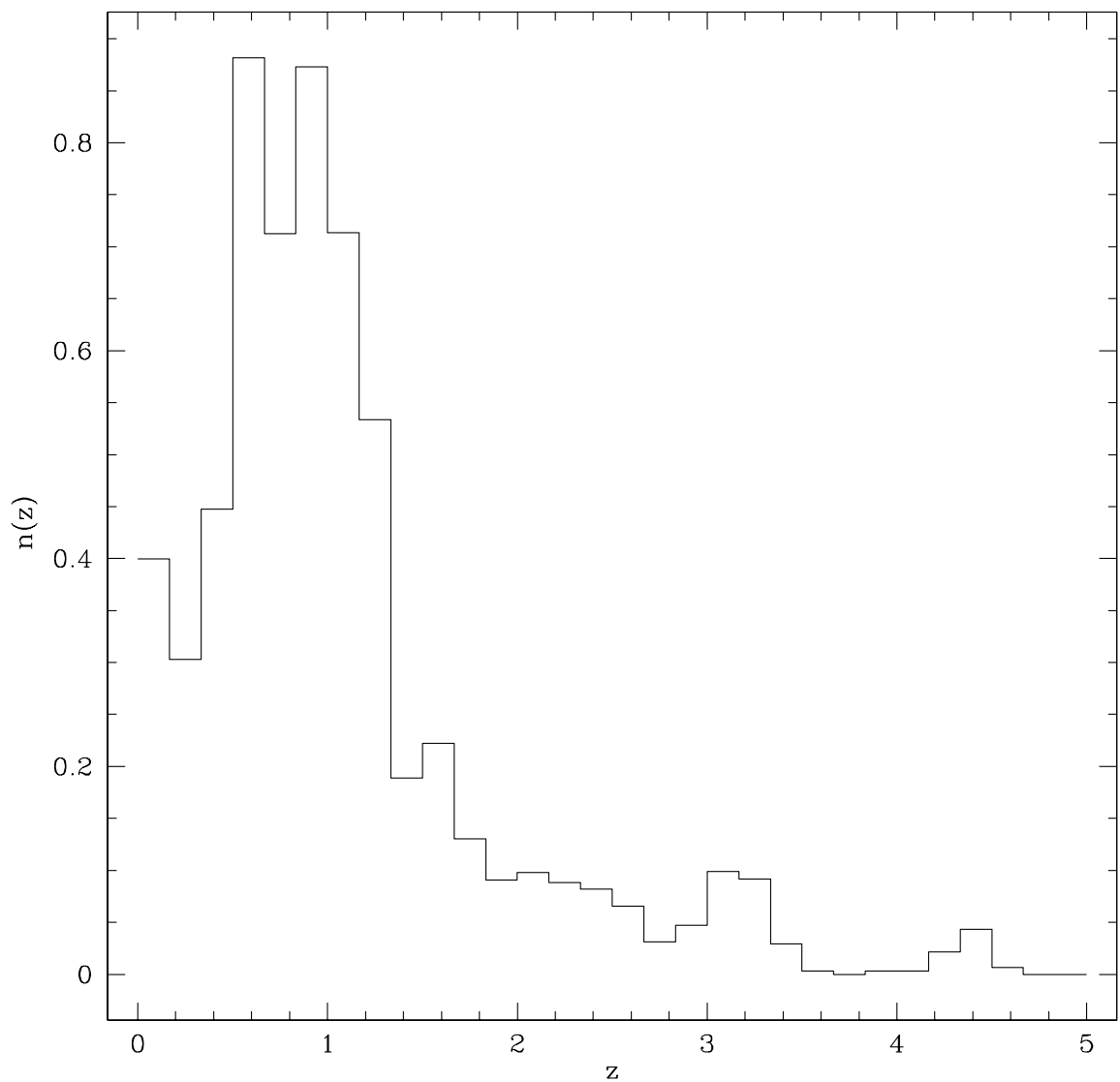}
\caption{The histogram shows the redshift distribution of the galaxies
with \protect{$I_{AB}>22$} with their appropriate noise weighting.
It was computed from the
photometric redshifts of the Hubble Deep Fields north and
sourth (see \citet{2001A&A...374..757V}).}
\label{fig:nz}
\end{figure}

The resulting histogram
for this sample taken from the HDF is shown in Figure \ref{fig:nz}.
In the absence of any spectroscopic survey deeper than
$I_{AB}=22$ this is the best redshift estimate at the
moment. In the future, any accurate cosmological
parameter interpretation from the skewness will require
precise determination of the source redshift.

\section{Simulations}
\subsection{N-body}

Both the variance and skewness arise from non-linear scales, and
simulations are required to calibrate theory
\citep{2000ApJ...537....1W,2000ApJ...530..547J}.  The past studies had
shown that Gaussian treatments are not accurate for error bar
estimates.  The sample variance depends on the actual survey geometry,
for which simulations are the only tractible approach to quantify
these effects.

We ran a series of N-body simulations with different values of
$\Omega_0$ to generate convergence maps and make simulated catalogs to
calibrate the observational data and estimate errors in the analysis.
The power spectra for given parameters were generated using CMBFAST
\citep{1996ApJ...469..437S} and these tabulated functions were used to
generate initial conditions.  The power spectra were normalized to be
consistent with the earlier two point analysis from this data set
\citep{2002A&A...393..369V}.  We ran all of the simulations using a
parallel, Particle-mesh N-body code (Dubinski, J., Kim, J., Park, C.
2003, in preparation) at $1024^3$ mesh resolution using $512^3$
particles on an 8 node quad processor Itanium Beowulf cluster at CITA.
Output times were determined by the appropriate tiling of the light
cone volume with joined co-moving boxes from $z \approx 6$ to $z=0$.
We output periodic surface density maps at $2048^2$ resolution along
the 3 independent directions of the cube at each output interval.
These maps represent the raw output for the run and are used to
generate convergence maps in the thin lens and Born approximations by
stacking the images with the appropriate weights through the comoving
volume contained in the past light cone.

All simulations started at an initial redshift $z_i=50$, and ran for
1000 steps in equal expansion factor ratios, with box size $L=200 h^{-1}$
Mpc comoving.  We adopted a Hubble constant $h=0.7$ and a scale invariant
$n=1$ initial power spectrum.  A flat cosmological model with  $\Omega_0 +
\Lambda = 1$ was used.  Four models were run, with $\Omega_0$ of
0.2, 0.3, 0.4 and 1.  The power spectrum normalization $\sigma_8$ was
chosen as 1.16, 0.90, 0.82 and 0.57 respectively.

We used the simulations in two modes.  To study scaling relations and
quick analyses, we projected them to convergence maps which were all
analyzed with idealized noise on a square domain.  We also used the
simulations to make mock catalogs which we processed through the full
pipeline that was used on the real data.  Using the mock catalogs we
can quantify the effect of sample variance on $S_3$.  The errors are
expected to be asymmetric: when $\Omega_0$ is large, skewness is small,
and the sample variance in skewness is also small.  So if one measures
a large skewness, one can strongly rule out a high value of $\Omega_0$.
But if intrinsic skewness is large, its sample variance is also large.
So measuring a small skewness does not rule out a low $\Omega_0$ unless
one has a very large field and good statistics.

\subsection{Simulated Convergence and Shear Maps}

Using our N-body simulations we generated two sets of maps.  One is a set of
idealized $\kappa$ convergence maps, from which we can easily measure
noise-free statistical quantities.  The second is a set of $\gamma$ shear maps
sampled at the actual 2-D galaxy positions of the survey, which are 
processed through the same
pipeline as the actual observed Virmos-Descart data set.  
The redshift weights were taken to be the statistical average, so
this does not take into account the source-lens-clustering effect
\citep{2002MNRAS.330..365H,1998A&A...338..375B}.
Comparing the
results from the two procedures allows us to cross-check our analysis,
and adds confidence to the interpretation of the complex statistical
procedure in terms of simulated viewable dark matter maps.

The convergence $\kappa$ is the projection of the matter over-density $\delta$
along the line of sight $\hat{\theta}$ weighted by the lensing
geometry and source galaxy distribution. It can be expressed as 
\begin{equation}
\kappa(\hat{\theta})=\frac{3}{2}\Omega_0\int_0^{\infty} W(z) 
\delta(\chi,\hat{\theta})d\chi.
\end{equation}
$\chi$
is the comoving distance in units of $c/H_0$. The Hubble constant
is parametrized by $H_0=100\  h 
\ {\rm km/s/Mpc}$. The weight
function $W(z)$ is determined by  the source galaxy distribution function
$n(z)$ and the lensing geometry. For a flat universe, 
\begin{equation}
W(z)=(1+z) \chi(z)\int_{z}^{\infty} n(z_s) [1-\chi(z)/\chi(z_s)] dz_s.
\end{equation}
$n(z)$ is normalized such that $\int_0^{\infty} n(z)dz=1$. 
For Virmos-Descart, we adopt the distribution shown in figure
\ref{fig:nz}, which is not easily fit by an analytic function.


During each simulation we store 2D projections to the midplane of the
dark matter overdensity $\delta \equiv \delta \rho/\bar{\rho}$ through
the 3D box at every light crossing time through the box along each of
the x, y and z axes.  Our 2D surface density sectional maps are stored
on $2048^2$ grids. After the simulation, we stack sectional maps
separated by the width of the simulation box, randomly choosing the
center of each section and randomly rotating and flipping each
section.  The periodic boundary condition guarantees that there are no
discontinuities in any of the maps.

\begin{figure}
\plotone{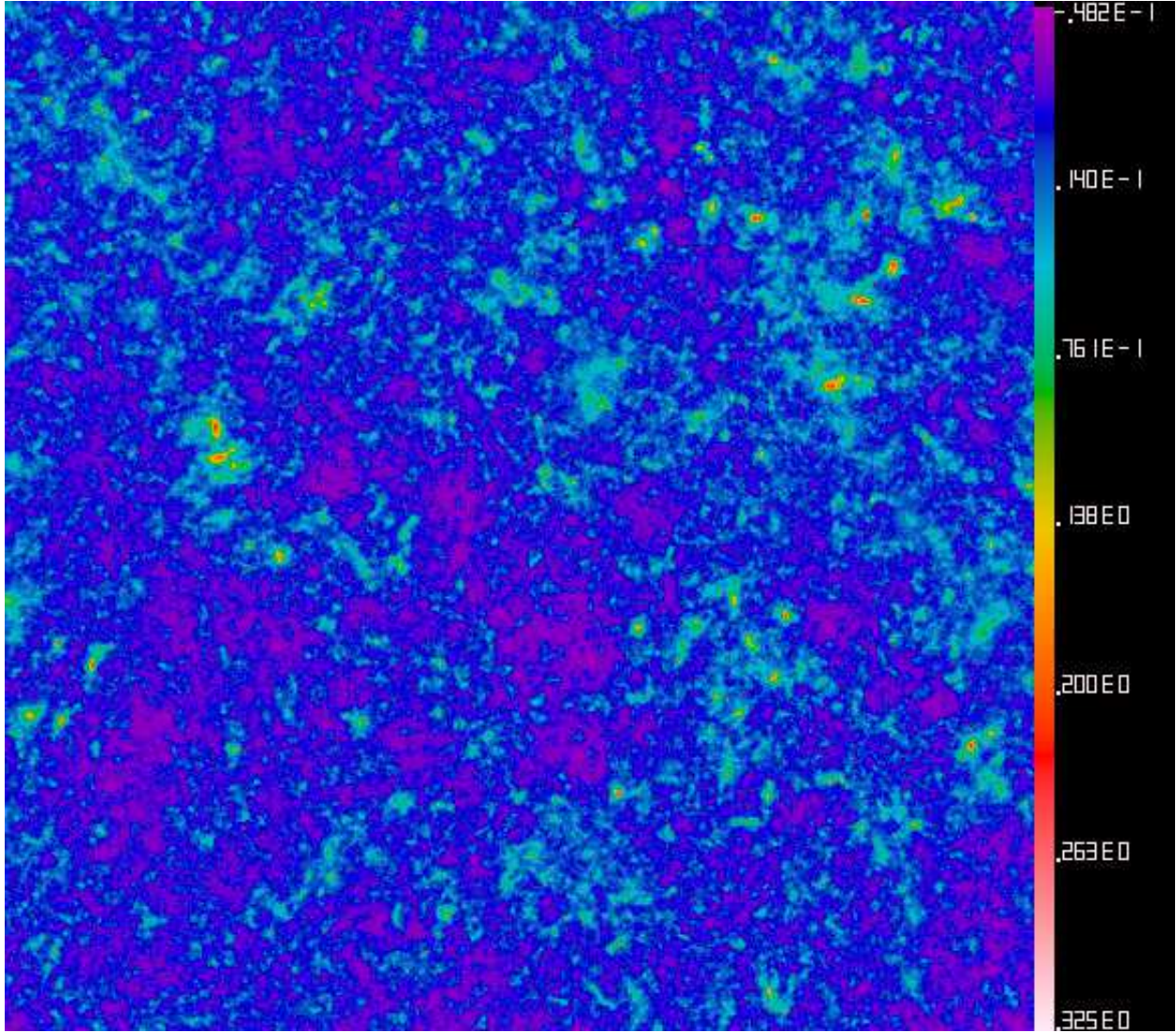}
\caption{A $\kappa$ map in the
N-body simulation of  a $\Omega_0=0.4$
$\Lambda$CDM cosmology. The map width is 3.18
degree and has $2048^2$ pixels. The scale is in units of $\kappa$.
The skewness of the distribution is apparent.  Decreasing the
cosmological density while maintaining the same variance forces
structures to become more linear, and thus more skewed.
\label{fig:lensingmap}} 
\end{figure}

We then add these sections with the weights given by $W(z)$ onto a map
of constant angular size, which is generally determined by the maximum
projection redshift.  To minimize the repetition of the same structures
in the projection, we alternatively choose the sectional maps of x, y,
z directions during the stacking. Using different random seeds for the
alignments and rotations, we make $40$ maps for each cosmology. Since the
galaxy distribution peaks at $z\sim 1$, the peak contribution of lensing
comes from $z\sim 0.5$ due to the lensing geometry term.  Thus a maximum
projection redshift $z\sim 2$ is sufficient for the lensing analysis.
We integrate the maps for the $\Omega_0=1,0.4,0.3$ simulations to $z=2$
and obtain $40$ maps each with angular width $\theta_\kappa=4.09, 3.18$
and $3.02$ degrees, respectively. To make maps for  $\Omega_0=0.2$ which
are large enough to fit the survey fields, we truncated the integral
at to $z=1.8$ and obtained maps with angular width $\theta_\kappa=2.86$
degree.  We show one $\kappa$ map from an $\Omega_0=0.4$ simulation in
fig. \ref{fig:lensingmap}.  At this resolution, the skewness is quite
obvious.  The challenge is to recover the skewness from a noisy
observation of such a data set.

From the sectional maps, we also produce shear maps. These $\gamma$
maps are used to produce mock galaxy catalogs to check the data
analysis pipeline and to calibrate the results. For each sectional
$\kappa$ maps, we computed sectional $\gamma_1$ and $\gamma_2$ maps
using FFT's.  The periodic boundary conditions on each section allow
us to circumvent the boundary condition issue.  Then by the same
projection procedure, we make $40$ $\gamma_1$ and $\gamma_2$ maps with
the same fields of view.  We sample these simulated shear maps at the
actual galaxy position from the Virmos-Descart catalog, which ensures
that the mock catalogs have same masks and weights.  Since the
Virmos-Descart catalog consists of four fields, we can make ten full
sets of mock catalogs from each simulation.

With these maps, we simulate the Virmos-Descart survey. Our goal is 
to measure $S_3$ of smoothed $\kappa$ fields and its dispersion and
distribution in 
simulated survey catalogs to provide calibrations for the data
analysis.  In this paper, we fix the smoothing function as the
compensated Gaussian function with a filter radius $\theta_f=5.374$ arc min.
The  maps we obtained above are non-periodic after the projection. In 
order to eliminate the edge effect,  we cut $10\%$ from the four margins 
of each smoothed $\kappa$ map. For the noise-free
maps, the $S_3$ parameter is defined as 
\begin{equation} 
\label{eqn:cleans3}
S_3(\theta_f)=\frac{\langle \bk^3 \rangle}{\langle \bk^2\rangle^2}.
\end{equation}
We used the $\kappa$ maps to fit the dependencies of $S_3$ on
$\Omega_0$.  We parametrized the relation as 
\begin{equation}
S_3=\alpha \Omega_0^{-\beta}.
\label{eqn:s3}
\end{equation}
The best fit values for $\alpha$ and $\beta$ are
\begin{equation}
\alpha=66.0\pm 2.3,\ \ \ \ \beta=-0.64\pm 0.03,
\end{equation}
which are similar to those obtained from second order perturbation theory
with different window functions.

\section{Analysis}
\subsection{Optimality}

The optimal estimation of power spectra for Gaussian random fields is
well understood.  It requires a weighting of the data by the inverse
of signal plus noise correlators.  This procedure generally costs of
order $O(N^3)$ where $N$ is the number of data points, about one million
for our catalog where we have two observables, $\gamma_1,\gamma_2$
per galaxy.  Clearly the optimal procedure is prohibitively expensive.
In principle rapid multi-scale iterative algorithms allow an analysis
in $O(N)$ \citep{2002astro.ph.10478P}.  The cost prefactor in that algorithm
is still large, and we decided to use the two point correlation
function instead, whose computational cost is $O(N\log N)$ with a
small prefactor.

Computing the two point correlation function is in general not an
optimal estimator of the power.  Being a quadratic estimator, it
weights each data bin by only its local noise.  Each correlation
function bin $q_i$ is an inverse noise $N^{-1}$ weighted estimator of the data
$\Delta$ for a bilinear form $Q_i$:
\begin{equation}
q^i = \Delta^T N^{-1} Q_i N^{-1} \Delta
\end{equation}
while an optimal estimator should have used $(S+N)^{-1}$ as its
weights instead.  In the limit that signal to noise is small, the
correlation function is optimal, similarly it is optimal if the signal and
noise covariance matrices commute.  For weak lensing surveys, on small
scales signal to 
noise is small, while on larger scales the coverage is reasonably
uniform, so a correlation function is not a very poor estimator.

To discuss the accuracy of the higher moments of a slightly
non-Gaussian random field, consider the one point statistic of a
filtered convergence field.  The intrinsic distribution of $\bk$ is
described by $P_\kappa$. We observe a noisy realization thereof,
$\bk^o=\bk+\nu$, with $\nu$ denotes a noise random variable.  Its
variance is $\sigma_o^2=\sigma_\bk^2+\sigma_\nu^2$.  Typically the
variance is noise dominated, in which limit the fractional accuracy of
measuring $\sigma_\bk^2$ when $\sigma_\nu^2$ is known is
\begin{equation}
\frac{\Delta \sigma_\bk^2}{\sigma_\bk^2}
=\sqrt{\frac{2}{n}}\frac{\sigma^2_\nu}{\sigma^2_\bk}
\label{eqn:v2}
\end{equation}
where $n$ is the number of independent samples.  On the scales of a
few arc minutes where this is best measured, signal to noise is below
unity, and an accurate measurement is done statistically
where one can reduce the noise due to the
large number of arc minute sized patches in the survey.

The skewness is even harder to measure.  One expects
\begin{equation}
\frac{\Delta \bk^3}{\bk^3}=\sqrt{\frac{15}{n}}
\left(\frac{\sigma_\nu}{\sigma_\bk}\right)^3 \frac{1}{S_3 \sigma_\bk}
\label{eqn:errskew}
\end{equation}
The last term is of order unity. Comparing with (\ref{eqn:v2}) we
notice three losses: 1. The numerical prefactor increases. 2. The
error scales as a higher power of signal to noise. 3. The denominator
has increased powers of the converge variance, which is itself a small
number.  These trends suggest that higher moments are increasingly
difficult to measure in noisy data.

Optimality of measuring the three point function is a challenging
problem.  Gaussian fields have zero three point function, so its
distribution is intrinsically non-linear and must be measured either
through high order perturbation theory or in simulations.  By
optimality of an estimator we mean one which minimizes the variance.
The variance of the three point function is related to the six point
function, which is a complex problem.

We thus scale back our ambitions, and concentrate on the skewness
in a smoothed density field.  In analogy with the two point function,
we expect that an inverse noise squared weighting should be a
reasonable procedure in this noise dominated measurement.

\subsection{Two point functions}

We first review the efficient computation of two point functions as a
warm up exercise for the three point.  Computationally, we start with the
two shear estimates for each galaxy, and its noise estimate.  The weight
for each galaxy is the inverse noise variance.  
The noise variance is the quadratic sum of the intrinsic ellipticity
variance $\sigma_i$, and the noise variance due to the PSF correction
$\sigma_e$. The computation
of $\sigma_e$ is described in \citet{2002A&A...393..369V}: the galaxy
catalogue is divided into cells of 30 galaxies each in the
size-magnitude space. For each cell, the r.m.s. of the ellipticity
correction gives an estimate of $\sigma_e$. We arbitrarily put a lower bound of
0.28 for the effective noise of any galaxy, as we do not want to give
too large a weight to any individual object.  Varying this cutoff
does not seem to have significant effects on the results.
The resulting weight per 
galaxy is therefore:
\begin{equation}
w_i={1\over {\rm max}(0.28,\sigma_e)^2}.
\end{equation}

Galaxies are mapped onto
a grid using the cloud-in-cell algorithm \citep{1988csup.book.....H}.
The same is done for the weights.  We now have three gridded quantities,
which we fast Fourier transform.  The autocorrelation of the weights
is the inverse Fourier transform of the weight grid times its complex
conjugate.  Similarly, we obtain the two autocorrelations of the shear
components, and two cross correlations, we denote them $\tilde{\xi}_{ij}$.
If the dimensions of the grid are a factor of two larger than the spatial
extent of the galaxies, the convolution on the grid does not introduce
spurious boundary conditions.

The correlation functions are then multiplied by the angular weights,
to yield two raw angular averaged correlation functions
\begin{eqnarray}
\tilde{\xi}_+(r)&=&\frac{1}{2\pi}\int
[\xi_{11}(r,\theta)+\xi_{22}(r,\theta)] d\theta 
\nonumber\\
\tilde{\xi}_-(r)&=&\frac{1}{2\pi}\int
[\xi_{11}(r,\theta)-\xi_{22}(r,\theta)]\cos(4\theta)
+[\xi_{12}(r,\theta)+\xi_{21}(r,\theta)] \sin(4 \theta) d\theta
\end{eqnarray}
and an angular averaged weight correlation $\tilde{\xi}_w$.  The final
correlation is then
\begin{equation}
\xi_{\pm}=\frac{\tilde{\xi}_\pm}{\tilde{\xi}_w}.
\end{equation}

This allows a rapid computation of the two point function in eight Fourier
transforms, each costing $ 2.5N\log_2(N)$ operations, where
$N=4n_xn_y$ are the dimensions of the discretized grid, and the factor
of 4 comes from having to use a Fourier grid of double the size to
deal with the non-periodic boundary conditions.  On a
PC with the optimized Intel IPP or MKL libraries, the machines can
sustain over a gigaflop on a 2 Ghz processor, so using even very
fine grids $n_x\sim 10,000$ takes a matter of seconds.

The variance of the noise is uncorrelated between bins of the two
point correlation function.  One can see that the covariance of
the two point function is a four point function.  For a Gaussian field,
that can be expanded in terms of two point functions.  When the noise
is white, a four point function only expands into a non-zero quantity
when the four points have two coincident pairs.  This is not possible
when cross correlating two different correlation lags, and the
covariance is thus zero. The error due to noise in each bin of the
correlation function is $4/\tilde{\xi}_w$.

In tensor notation, the two point function can be written as a product
of functions which depend on the magnitude of separation $r$, and
angles of the 
unit separation vector $\hat{\bx}$
\begin{eqnarray}
\xi_{abcd}(\bx) &\equiv& \langle \gamma_{ab}(0)
\gamma_{cd}(\bx)\rangle 
\nonumber \\
&=&
\frac{\xi_+(r)-\xi_-(r)}{2}
(\delta_{ac}\delta_{bd}+\delta_{ad}\delta_{bc}-\delta_{ab}\delta_{cd})
+4\xi_-(r)\left(\hat{x}_a\hat{x}_b-\frac{\delta_{ab}}{2}\right)
\left(\hat{x}_c\hat{x}_d-\frac{\delta_{cd}}{2}\right)
\label{eqn:tshear}
\end{eqnarray}

\subsection{Three point functions}

The three point function is computationally  more
complex.  We describe here the algorithm we implemented which
computes it in $O(N^2\log N)$ operations for $N$ galaxies.
We first describe the mathematical enumeration of the three point
function which is prohibitively expensive.  We then describe a faster
procedure which computes the same quantity in a shorter time.

At every point we measure two components of the shear.  The shear
three point function has eight components.  To reduce the complexity
of the problem, we express the components of the shear in the
coordinate system of the line connecting the first two points.  We
define the first point to be at the origin, and the second point along
the x-axis at some distance $r$.  The third point can be anywhere in
the upper half plane.  For each galaxy, we first label it point 1.  We
search rightward, and consider the set of all pairs formed from point
1 and a rightward neighbor, which we successively label as point 2.
For each pair 1-2, we consider all galaxies with x coordinates larger
than that of galaxy 1, and y coordinates above the line connecting
point 1 to point 2.  This results in a unique counting of triangles.
As for the two point function, we multiply the shear estimates by the
weight.  We then compute the raw three point functions for the
weighted shear, as well as for the weights.  The final three point
function is the quotient of the two.

At face value this would appear to be an $O(N^3)$ operation, which is
prohibitively expensive for $\sim 10^5$ galaxies which we currently
have in each field.  Analogous to particle-mesh simulations, we map
all galaxies onto a two dimensional grid using a chaining mesh.  This
results in a linked list of galaxies residing in each cell, and costs
$O(N)$ operations to construct.  This allows us to find galaxies within
some locus without searching the whole list.

We enumerate all pairs of galaxies.  Each pair has a separation
and an angle of the line connecting them relative to the x-axis.  We
grid the list of all pairs onto a four dimensional grid labeled by the
position of the first galaxy, and their separations and angles.  The
grid of separations are chosen logarithmically.  The angles run in the
range $(-\pi/2,\pi/2)$, since we only need to count each pair once.
For each separation and angle, we convolve this list of pairs with the
gridded field of galaxies.  We accumulate the three point function by
averaging over the angles.  In this process, we use only the
pair-singlet cross correlation at lags with $\Delta x>0$ and $y$
values above the line describing the first pair.  This way all
triangles are counted exactly once.

In practice, gridding a Virmos-Descart field with 100,000 galaxies on
$512^2$ cells, using 40 logarithmic separations and 20 angles takes
about 15 minutes on the CITA GS320 alpha server which has 32 alpha
processors running at 731 Mhz.  The procedure requires 15GB of memory
to store the four dimensional list of pairs.

The final three point function is defined on a three dimensional grid.
In the intermediate pair enumeration we used polar coordinates to save
storage and reduce computational cost.  Since each pair is tagged by
the coordinate of the left galaxy, the three point function has some
angular smearing (in our case 9 degrees), so that large relative
errors can accrue when galaxies 2 and 3 are nearby to each other but
far from galaxy 1.  This can be overcome by mapping $x$ to $-x$ and
repeating the procedure.  For the windowed skewnesses these skinny
triangles do not contribute much, and we neglect this effect.  In
principle one could construct a three point function in $O(N \log(N))$
using a tree algorithm.  The challenge is that the opening angle
has to be chosen small since the shear needs to be rotated by the mean
angle.

\subsection{Skewness}
We review the integration of the third moment from the three point
function. 
A smoothed $\kappa$ map is defined as
\begin{equation}
\bk(\bx)=\int \kappa(\bx') \calu(|\bx-\bx'|) d^2 x'
\end{equation}
so the third moment can be expressed in terms of an integral over
the three point function
\begin{eqnarray}
\langle \bk^3(\bx)\rangle&=&\frac{1}{A}\int 
\kappa(\bx')\kappa(\bx'')\kappa(\bx''')
 \calu(|\bx-\bx'|)\calu(|\bx-\bx''|)\calu(|\bx-\bx'''|) 
d^2x d^2x' d^2x'' d^2x''' \\
&=& 2 \pi \int \xi_3(r,\bx) U_3(r,\bx) r dr d^2x
\label{eqn:3ptmom}
\end{eqnarray}
where $A$ is the area of integration.
The three point function $\xi_3$ is defined with the first point at the 
origin, the second along the x axis, and the third anywhere in the
plane.  The window $U_3$ is the overlap integral of three filter
functions placed at the same three locations. 
Due to symmetry, it is sufficient to consider only the upper half of the 
plane.

The integrands can be evaluated analytically for a compensated
Gaussian filter \citep{2002ApJ...568...20C}.  We choose
\begin{equation}
\calu=\frac{1}{2\pi r_0^2}\left(1-\frac{r^2}{2r_0^2}\right)
\exp\left(\frac{ -r^2}{2 r_0^2}+1\right).
\label{eqn:calu}
\end{equation}
This filter has zero area, and is normalized to have a peak amplitude
of unity in Fourier space.  This has the feature that the filter will
only damp modes, and never amplify.  Its Fourier transform is
$\calu(k)=(k r_0)^2 \exp(-k^2 r_0^2/2+1)/2$.  The filter peaks at wave
number $k_0=\sqrt{2}/r_0$.  When $r_0$ is measured in radians, $k$ is
the spherical harmonic number $l$.  When integrated over a flat
spectrum with constant $l^2 C_l$, the area in Fourier space is
$e^2/8$.  One can use this window as a reasonable broad band Fourier
power estimator.  One obtains the equivalent flat power by multiplying
the variance by $8/e^2$.  The effective full width at half max
width of the filter in
Fourier space is a multiplicative factor of 2.34.  

The corresponding shear filter is
\begin{equation}
\calq=\calu-\frac{2}{r^2}\int_0^r r' \calu(r') dr' = -\frac{r^2}{
4\pi r_0^4} \exp\left(\frac{ -r^2}{2 r_0^2}+1\right)
\end{equation}
The filtered convergence field is then
\begin{equation}
\bk(\bx)=\int \gamma_{ij}(\bx') \calq(|\bx-\bx'|)
\Delta\hat{x}_i \Delta\hat{x}_j d^2 x'.
\end{equation}
Due to the $r^2$ term in front of $\calq$, the hatted objects in the
integral are polynomials, and the integrand contains only Gaussian
integrals over polynomials, which are exactly solvable.  The smoothed
variance can be derived in terms of the tensor shear correlation
(\ref{eqn:tshear})
\begin{equation}
\langle\bk_E^2\rangle = \int \xi_{abcd}(\bx) W_{abcd}(\bx)d^2x
\end{equation}
for the window function
\begin{eqnarray}
W_{abcd}(\bx)&=&\frac{1}{4(2\pi)^2r_0^4}
\int x'_ax'_b(x'_c-x_c)(x'_d-x_d) 
\exp\left[-\frac{x'^2+(x'-x)^2}{2r_0^2}+2\right] d^2x'
\nonumber \\
&=& \frac{4(\delta_{ac}\delta_{bd}+\delta_{ad}\delta_{bc})
r_0^4-2(\delta_{ac}x_bx_d+\delta_{ad}x_bx_c+\delta_{bc}x_ax_d
+\delta_{bd}x_ax_c)r_0^2+x_ax_bx_cx_d}{2^8 r_0^4 \pi}
\nonumber\\
&&\times \exp\left(-\frac{x^2}{4r_0^2}+2\right)
\label{eqn:wabcd}
\end{eqnarray}
where we have dropped all trace terms which do not contribute.

For the skewness we can similarly write
\begin{equation}
\langle \bk^3\rangle= 2 \pi \int \xi_{abcdef}(r,\bx) 
W_{abcdef}(r,\bx) r dr d^2x
\end{equation}
where the corresponding window function $W_{abcdef}$ is
an integral described below.

Since the filter (\ref{eqn:calu}) is the Laplacian of a Gaussian,
convolving by it is equivalent to first taking the Laplacian of the
$\kappa$ field, and convolving that by a Gaussian.  The Laplacian of
the $\kappa$ field is also a local second derivative of the shear
field, for which a purely local decomposition of the shear field into
$E$ and $B$ mode results independent of survey geometry.  We thus expect
minimal coupling between $E$ and $B$ modes from survey geometry.

\begin{figure}
\plotone{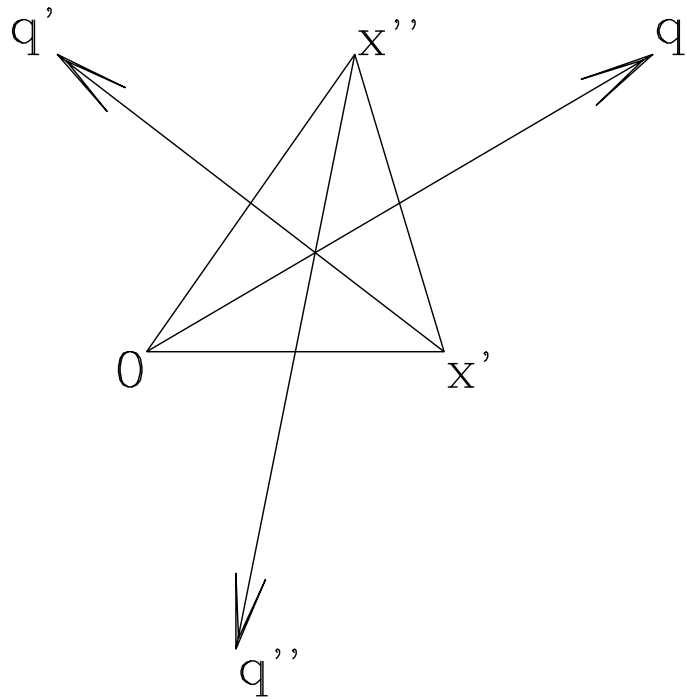}
\caption{three point configuration: we place one vertex of the
triangle at the origin, the second along the x axis, and the third in the
upper half plane.  The \protect{$q$} vectors are the axes along which
each shear component needs to be projected.}
\label{fig:triangle}
\end{figure}

To construct the analogue for the three point function,
it is convenient to define the vectors which point from each vertex to
the sum of the other two vertex vectors $q=x'+x''$, $q'=x''-2x'$ and
$q''=x'-2x''$.  They are depicted geometrically in Figure
\ref{fig:triangle}.

The non-vanishing terms in the integral are
\begin{eqnarray}
W_{abcdef}&=& \frac{e^3}{(2\pi)^2 2^3 3^7 r_0^{10}}
\left[q_a q_b q'_c q'_d q''_eq''_f \right. \nonumber\\
&& +3\times 4\delta_{ac}q_bq'_dq''_eq''_f
+({\rm 3\ cyclic\ permutations})\nonumber \\
&&+9(\delta_{ac}\delta_{be}+\delta_{ae}\delta_{bc})q'_dq''_f
+({\rm 12\ permutations})\nonumber \\
&&\left. +9\times 2(\delta_{ac}\delta_{bd}+\delta_{ad}\delta_{bc})q''_eq''_f
+({\rm 3\ cyclic\ permutations})\right]\nonumber \\
&&\times \exp\left[-\frac{x'^2+x''^2+(x''-x')^2}{6}\right]
\label{eqn:win3}
\end{eqnarray}
To check for systematics, one can compute the moments of the $B$ mode
skewness $\langle \bk_B^3\rangle$ by a 45 degree rotation,
corresponding to 
$(\gamma_1,\gamma_2) \longrightarrow (\gamma_2,-\gamma_1)$.  Similarly
one can get the moments $\langle \bk_E\bk_B^2\rangle$ and $\langle
\bk_E^2\bk_B\rangle$. When we substitute the three point window
(\ref{eqn:win3}) into three point integral (\ref{eqn:3ptmom}), we
place $\bx'$ along the x-axis, requiring only a one-dimensional
integral, and only integrate $\bx''$ on the upper half of the plane,
since we only count each triangle once.  The sum over the 6 indeces
only has 8 independent components, corresponding to the two shear
components at each of the three vertices.  


In analogy to the two point function, we can compute the expected
noise variance in each windowed skewness.  The covariance between bins
is again zero for white noise, since the only non-trivial six point
function consists of three pairs.  As for the two point function, we
measure the noise from the variance between bins in the three point
function, and assume that the noise is linearly proportional to the
inverse square root.  We verified the procedure by sampling four of the
simulated images in the $\Omega_0=0.2$ simulation on a regular $400^2$
grid with equal weights, and computing the skewness from the three
point function.  For these 2.86 degree images this skewness agreed
with the image skewness to better than ten percent on the 5.37 arc
minute scale, which is consistent with the variation expected from the
different weighting of the boundary.

\section{Results}

\begin{figure}
\plotone{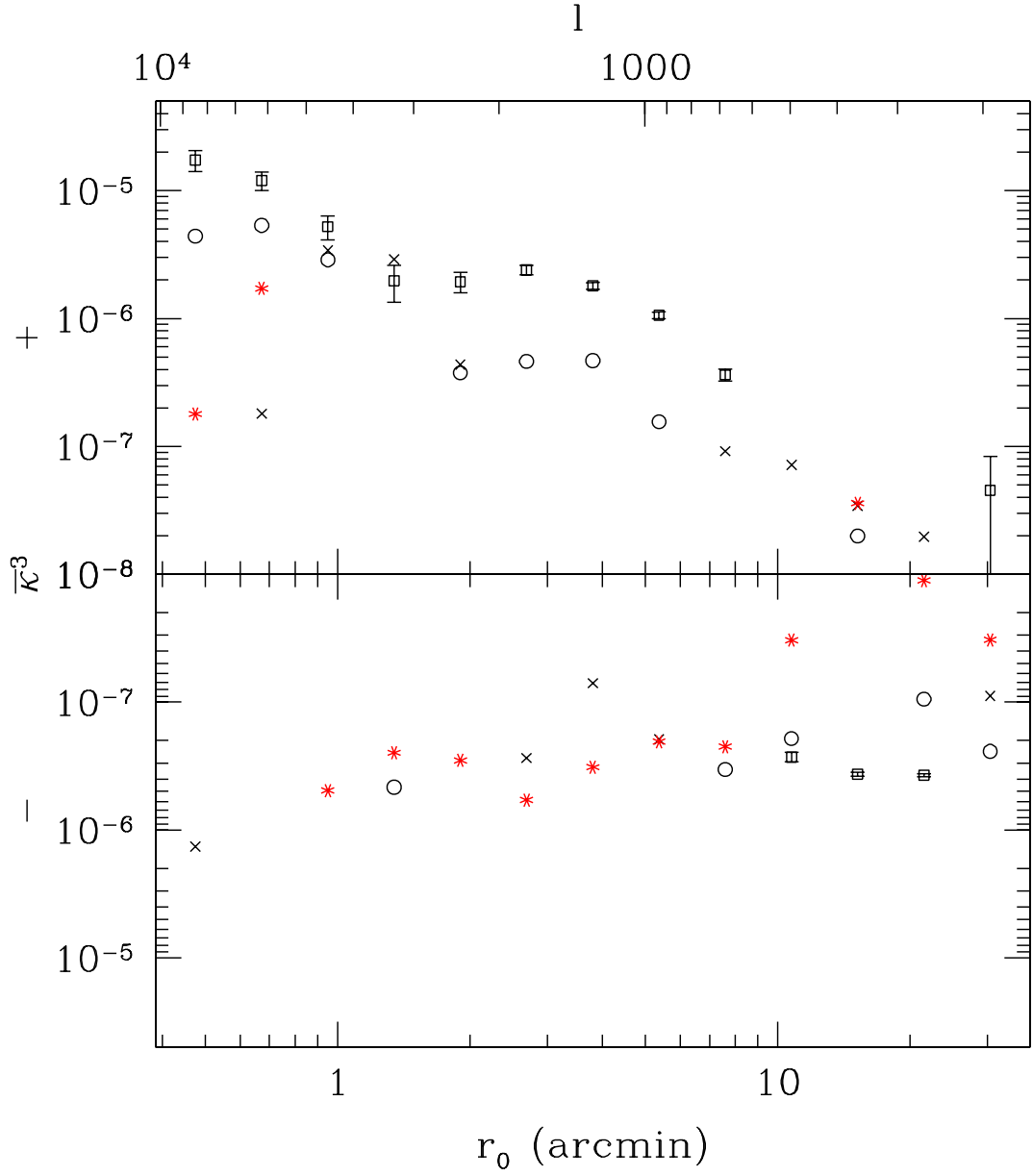}
\caption{
Skewness of the shear field. The lower panel shows the negative values
for any variable which goes below zero.
The boxes with error bars are the $E$ mode.
The stars are the $B$ mode, the circles are \protect{$EB^2$} and the
crosses are \protect{$E^2B$}.  The error bars reflect the scatter
expected in the absence of 
a signal, and does not include signal sample variance.
}
\label{fig:skew}
\end{figure}

\begin{figure}
\plotone{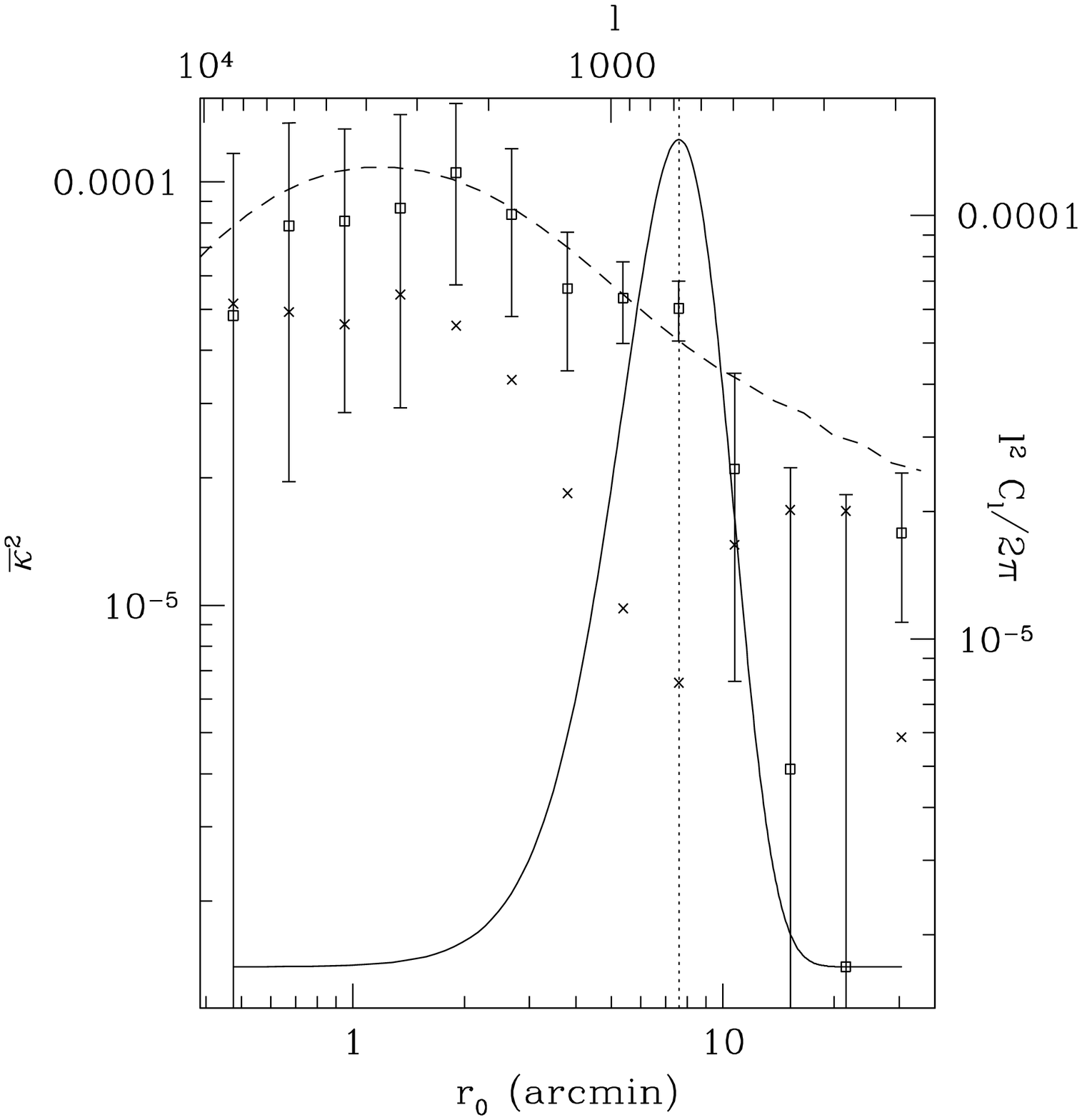}
\caption{
Variance of the shear field. The right side label is offset by
a factor of \protect{$8/e^2$} which  gives the
conversion to flat bandpower.
The boxes are the lensing \protect{$E$}
mode minus the non-lensing \protect{$B$} mode, while the crosses are
the non-lensing \protect{$B$} mode.  The error bars include the
statistical error from shot noise plus the amplitude of 
\protect{$B$}.  The
solid curve is the log-linear Fourier space window function used for
the current analysis,
which corresponds to a scale of 5.37 arc minutes.  The vertical dotted
line marks this scale.  The dashed curve is the ensemble average
obtained from 40 simulated maps in the $\Omega_0=0.4$ model.}
\label{fig:var}
\end{figure}

We show the measured skewness in Figure \ref{fig:skew}.  The variance
in the same filter from the two point function is shown in Figure
\ref{fig:var}.  We find that statistical errors from galaxy noise are
small, but on all scales we see some $B$ mode.  The $B$ mode is more
significant in the variance measurement than in the skewness.  This is
to be expected, since variances are always positive, while skewness
may cancel.

After having obtained these statistics, we were primarily concerned by
systematic errors, of which the $B$ mode is a potential diagnostic.
In our arbitrary binning, the scale at 5.37 arc minutes had the
smallest $B$ mode in the variance, so we proceeded with all further
analyses using this scale.  At this scale we find $ \bk^3 = 1.06\pm
0.06\times 10^{-6} $.  These errors are Gaussian and contain only
contributions from noise.  Similarly, we find a variance at the same
scale of $\bk^2= 6.30\pm 0.62\times 10^{-5}$.  To model the possible
effects of the $B$ mode, we added the observed $B$ power to the error
bar, and subtracted the measured $B$ power from the signal.  Our
adopted value for subsequent analysis is $\bk^2= 5.32\pm 0.62\pm 0.98
\times 10^{-5}$, where second error bar is the magnitude of the $B$
mode.

The B-mode may or may not contribute to the skewness, which depends
on the parity of the source of B modes.  
The measured value is consistent
with zero, so we neglect it in the skewness interpretation.
Since variances are always positive, one expects any spurious source
of systematics to increase the observed variance.  Any corrections due to
a observed $B$ mode would also correct the observed signal down.  For
the skewness, the effect could have either sign.  We therefore did not
correct the $\langle \bar{\kappa}^3\rangle$ values.

\begin{figure}
\plotone{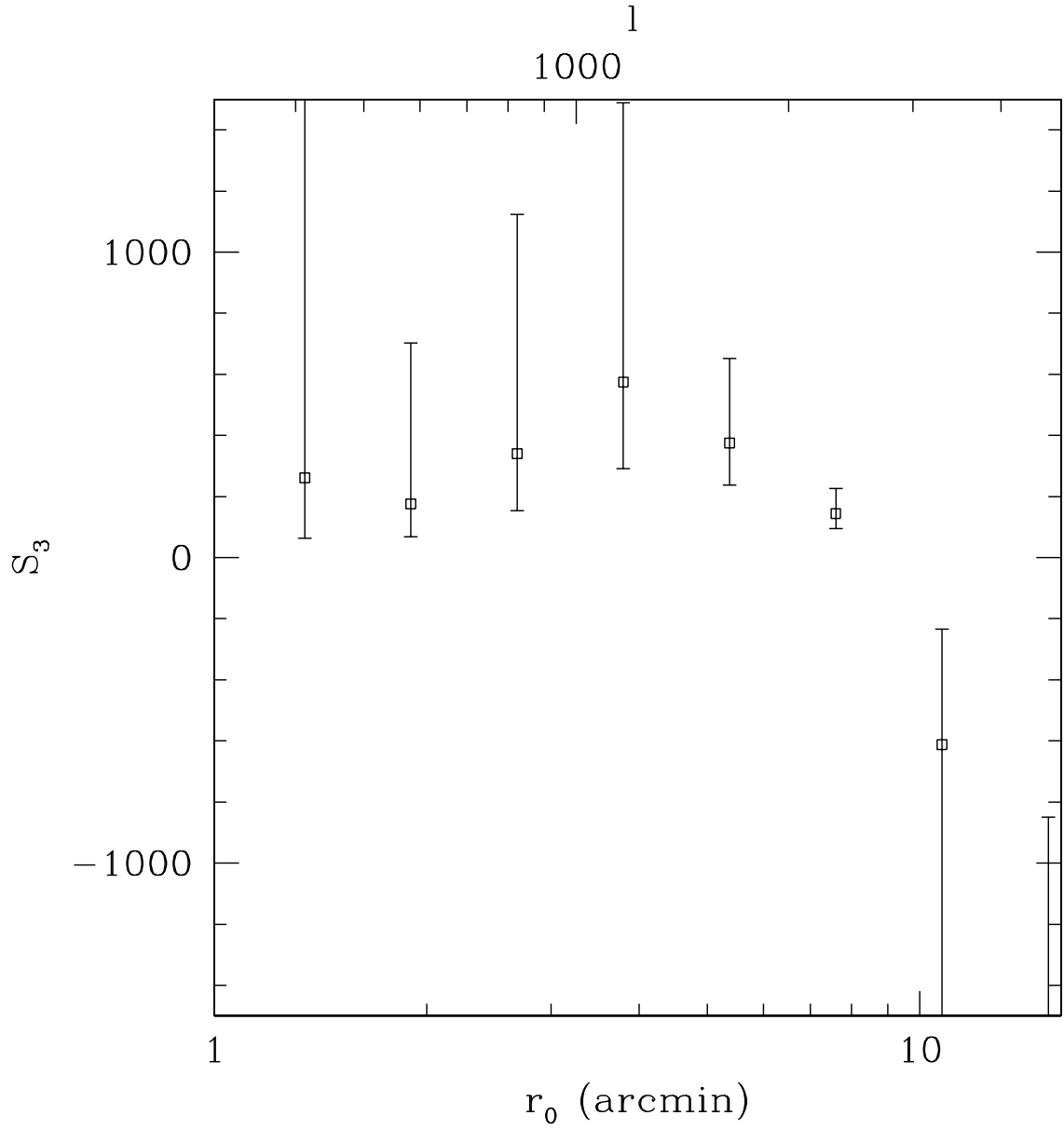}
\caption{
\protect{$S_3$} for the data, where we used \protect{$E^2-B^2$} as the
estimator for \protect{$\bk^2$}.  
We added the amplitude of the $B$ mode to the error bar, which is described
in the text.}
\label{fig:s3}
\end{figure}

Using the skewness and variances we can construct $S_3$.    The actual
value of the $B$ variance is added to to the error bar of the
variance.  
The noise errors
are Gaussian since they arise from the mean of a very large number
of data points.  In the absence of a better model, we also treat the
systematic error as Gaussian.  The resulting $S_3$ is shown in Figure
\ref{fig:s3}.  Random variables which are quotients of Gaussians, such as
$S_3$, don't generally have well defined moments.  The error bars in the
plot are obtained by taking one sigma on each of $\langle \bk^3 \rangle$
and $\langle \bk^2\rangle$, and plotting the observed change in $S_3$.
For sums of Gaussians this is like summing the standard deviations, which
is always an overestimate of the actual standard deviation of the sum.
One can consider the error bars in 
figure \ref{fig:s3} to be a conservative
estimate.  For our scale of 5.37 arc minutes, we used a second error
metric by Monte-Carlo sampling of Gaussian noise.  Nominally, $S_3=375$,
and the Monte-Carlo sample finds the median and percentile spreads are
$S_3=376^{+242}_{-125}$ at $68\%$ confidence, and $S_3=376^{+815}_{-194}$
at $95\%$.  Again, these error bars do not include sample variance,
which depends on cosmological models and require N-body simulations
to calibrate.

In principle, the fit ($\ref{eqn:s3}$) gives us a value of $\Omega_0$,
which would be a very small value, indicating that the data prefers a
low density universe with no practical lower bound.  The error bar and
confidence interval is a tricky task, since it requires knowledge of
the area and geometry of the sky patch surveyed.  In principle, it
should scale as the inverse square root of the number of independent
patches as given by Equation (\ref{eqn:errskew}).  At the scales of
the filter size of 5.37 arc minutes the survey is quite irregular, and
it is not easy to quantify the effective area of sky observed, nor the
correlation between patches. So to measure the scatter we use the
noise-free mock catalogs from simulations.  We process these catalogs
through the same three point function simulation pipeline.  Even
though the simulated catalogs have no noise, we used the same weights
as the noisy data.  This allows us to separate the error from sample
variance from the error from noise.

\begin{figure}
\plotone{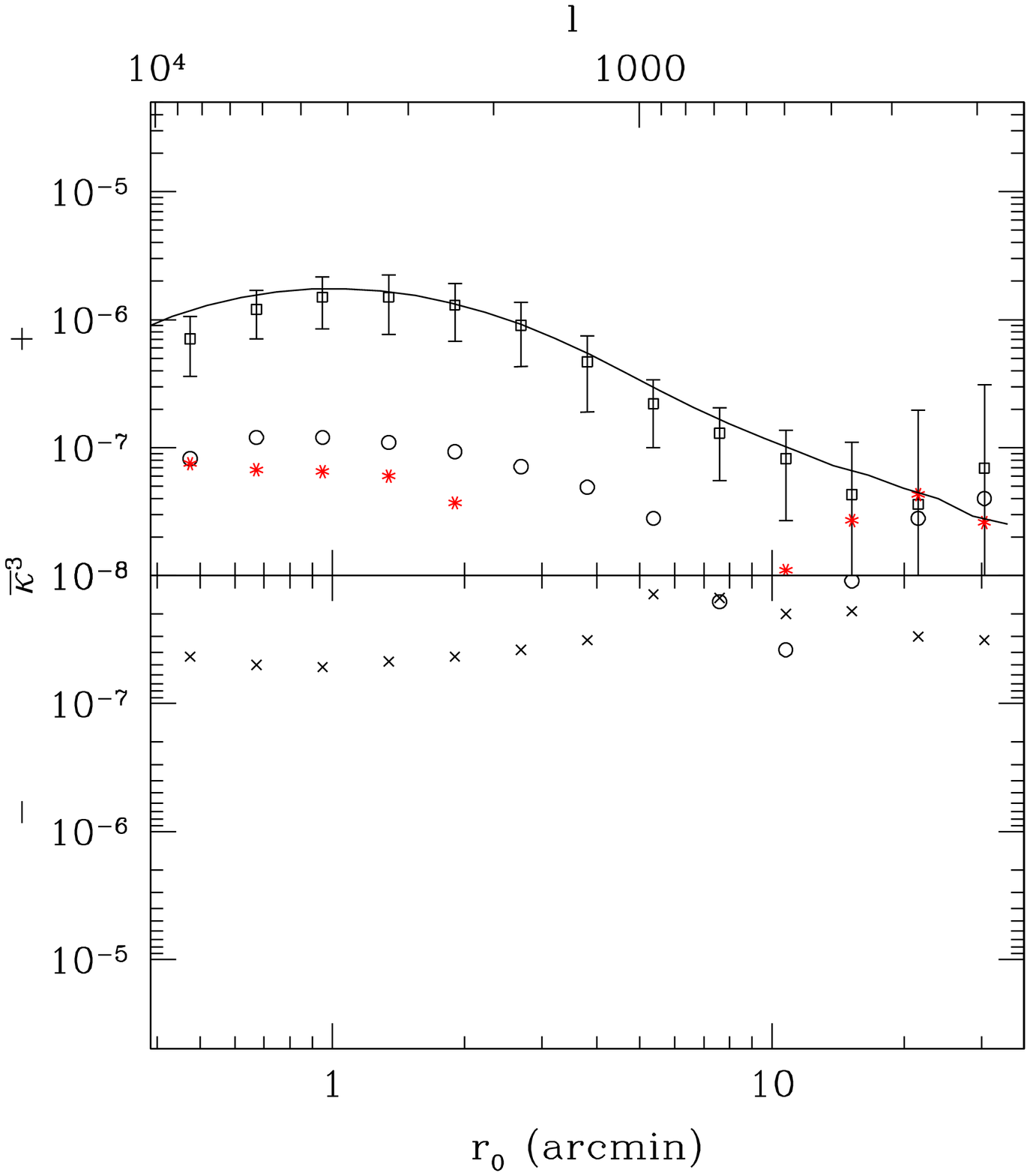}
\caption{Skewness in a noise free $\Omega_0=0.4$ simulation processed
through ten mock catalogs and the entire pipeline.  Actual galaxy positions
and weights were used.  The symbols have the same meaning as in figure 
\ref{fig:skew}, with the points denoting the average over ten mock catalogs,
and the error bar the standard deviation. The solid curve is the 
ensemble average
from the $\kappa$ maps.  The agreement is a cross check on the
analysis pipeline.  We note that the estimate of the standard
deviation has an error of 47\% with the 10 mock catalogues that were used.
}
\label{fig:simk3}
\end{figure}

\begin{figure}
\plotone{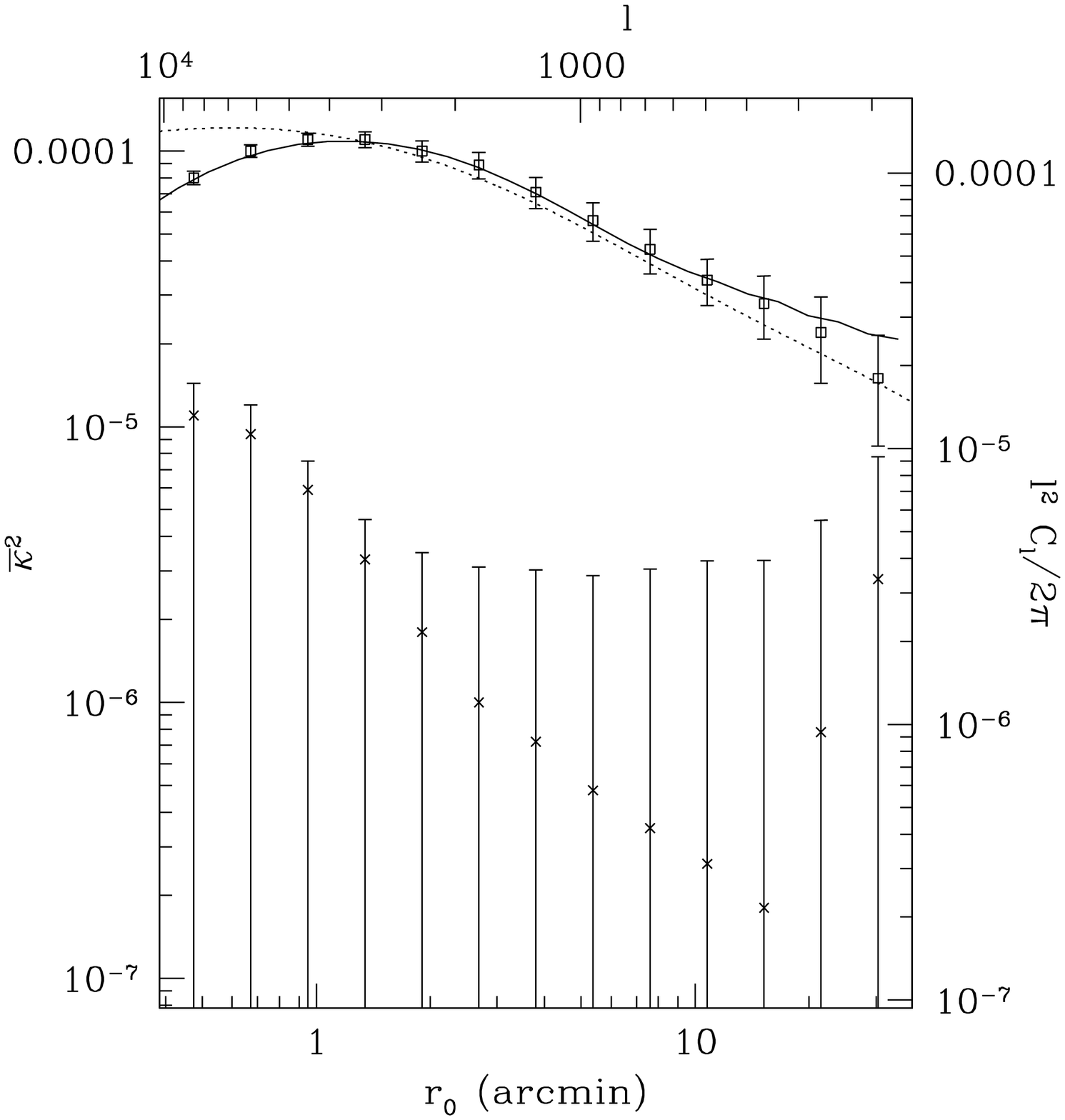}
\caption{Variances in the same simulation as figure \protect{\ref{fig:simk3}}.
The mode separation works better for the variance than for the skewness,
and the mock catalogs have significantly smaller sample variance on
$\bk^2$ than for $\bk^3$.  The solid curve is the same as the dashed
line in figure \protect\ref{fig:var}, which is the ensemble average
from the $\kappa$ maps.   The dotted line is the fit from \citet{1996MNRAS.280L..19P}.
}
\label{fig:simk2}
\end{figure}

In figures \ref{fig:simk3} and \ref{fig:simk2} we show the results
for a $\Omega_0=0.4$ model obtained through the pipeline.  The points
are the average from ten mock catalogs, and the error bars are the
standard deviation across the ten catalogs.  The mode separation for
the variances is typically better than 1\%, while skewness estimates
have crosstalk at the 10\% level.  The plots also include the ensemble
averages from the 40 direct idealized $\bk$ maps, processed through an
independent analysis which shares only the N-body simulation output,
which is a useful cross-check of the pipeline.  The fit from
\citet{1996MNRAS.280L..19P} is also plotted as the dotted line,
which provides a second reference for the resolution of the simulation.
We note good agreement for scales larger than 1 arc minute, but below
which the simulations are limited by resolution.

While the error bars in figures \ref{fig:skew} and \ref{fig:var} only
accounted for the noise, the simulation plots show only the sample
variance.  The sample variance is clearly decreasing at smaller
scales, where more independent patches are used.  At scales below one
arc minute, the simulations are probably limited by resolution. 

To estimate the sample variance errors, we used a sample of forty
simulated galaxy catalogs.  This number is limited by the expense of
computations.  The sample consisted of our four simulation models with
ten catalogs for each model.  We scaled the skewness and variance in
each model by the inverse of the standard deviation of the same
quantity obtained from the simulated maps.  This set of normalized
samples was used to Monte Carlo sample 10000 instances of $\bk^3$ and
$\bk^2$.  For a given trial value of $\Omega_0$, we fixed $\bk^2$ to
the observed value, and used $\bk^3$ from Equation \ref{eqn:s3}.  The
normalized samples are scaled to these constraints, and we randomly
draw from the 40 simulated values.  We add noise drawn from a Gaussian
with the appropriate variance.  This allows us to form a value for
$S_3$ for each sample.  We find the confidence bounds using a uniform
Bayesian prior. For $\Omega_0=0.25$ we found 20\% of the samples to
have $S_3$ larger than the observed value, while for $\Omega_0=0.5$
only 10\% of the samples hasd $S_3$ larger than observed.  For
$\Omega_0=1$ such large values of $S_3$ were never found.  We note
that our limited simulation sample only provides a crude estimate of
the errors.  It is probably not meaningful to derive 95\% statistics
from this small sample whose wings are non-Gaussian.  The preliminary
interpretation is that a low value of $\Omega_0$ is favored.  More
detailed systematic numerical studies are in progress.  No practical
lower bound on $\Omega_0$ is derived, since below $\Omega_0=0.2$ the
empirical fit from Equation (\ref{eqn:s3}) has not been calibrated.

We compared the sample variance in the simulated maps to that of the
mock catalogs.  For the scale of interest, we found the scatter in
$\bk^2$ from the mock catalogs shown in figure \ref{fig:simk2}
consistent with that of a 8.5 sq degree $\bk$ map, which is the
effective area sampled by our survey.  The scatter in $\bk^3$ from the
mock catalogs was significantly larger than in a 8.5 sq degree
contiguous map, possibly closer to what one sees across 3 sq degree maps.
This suggests that the inverse noise variance weights are not optimal
to estimate skewness.  Unfortunately with the limited number of mock
catalogs, the error on the sample variance is large, and more
simulations are needed.  To test the resolution of the simulations, we
also compared the results for a simulation with half as many particles
in directions on a grid twice as coarse.  In the $\Omega_0=0.3$ model,
the resulting $S_3$ on the $\bk$ maps changed by less than one
percent. The variance and skewness separately changed by about 10\%
and 20\% respectively, and one would expect the changes to be
significantly smaller compared to a higher resolution simulatio.  We
thus expect that simulation resolution should not be a major source of
error on these scales.

We also examined the variance between the four Virmos-Descart fields.  At the
scale of 5.37 arc minutes, the fields F14, F02, F22 and F10 had
$\bk^3$ of 
$8.7\times 10^{-8},
1.1\times 10^{-6},
-5.1\times 10^{-7},
1.1\times 10^{-6}$
respectively.  Their individual values of $\bk^2$
were 
$3.0\times 10^{-5},
7.0\times 10^{-5},
7.5\times 10^{-5},
7.0\times 10^{-5}$ without any subtraction of a $B$ mode.
This scatter is consistent with that seen in the mock catalogs.  
One could in principle use the observed scatter to model the sample
variance, but with only four fields, the error in the scatter is
$\sqrt{2/3}$ which is very large, so we used the larger sample of
simulated mock catalogs.

Several other factors also affect the skewness interpretation.
Source-Lens-Clustering \citep{2002MNRAS.330..365H,1998A&A...338..375B} 
for example systematically reduces
the observed skewness, which would strengthen the inferred bounds on
$\Omega_0$.  Using Table 3 from \citet{2002MNRAS.330..365H}, 
the closest model to
our data is probably C1, which for a non-evolving source distribution
and the closest comparable window size of 10 arc minutes results in a
16\% effect.

The current analysis is limited by sample variance, which will improve
dramatically with the CFHT legacy survey.  The error on the error
estimate is not yet well determined, requiring more simulations to
quantify.  Errors in the source redshift distribution also effect
the results.  These effects are expected to be small compared to
the sample variance.
Work on this issue is in progress, and we expect an order
of magnitude improvement in the near future.  The precise numerical
computation of skewness is also a challenge, as different groups
measure slightly different normalizations for the
variance \citep{2000ApJ...537....1W,2000ApJ...530..547J}.  For the
current data, the differences are not substantial, but need to be
tightened with larger and more accurate galaxy catalogs.  Currently
only one angular scale was used, which is one with the smallest 
$B$ mode contamination in the variance measurement.  Even on this
optimal scale, the $B$ mode will soon become a limiting factor when
the survey area is increased.

\section{Conclusions}

We have presented the first direct measurement of dark matter skewness
using weak gravitational lensing in the Virmos-Descart survey.
We chose the compensated Gaussian with a scale of 5.37 arc minutes for
all analyses, which had the smallest $B$ mode, consistent with zero.
At this scale, the skewness of the dark matter is $\bk^3= 1.06\pm 0.06
\times 10^{-6}$.  We measured the variance $\bk^2= 5.32\pm 0.62\pm 0.98
\times 10^{-5}$.

This implies $S_3=375^{+342}_{-124}$.  Calibrating to mock catalogs
from N-body simulations, we find $\Omega_0<0.5$ at 90\% and
$\Omega_0<0.25$ at 80\% confidence.  The errors are dominated by sample
variance.  The CFHT legacy survey will increase the sky area by a
factor of 20, which should result in a decrease of a factor of 4 in
the error in a year's time.  Better constraints could also be obtained
by going to smaller scales, where sample variance is smaller.  This
requires higher resolution simulations, faster three point analysis,
and a better understanding of the origin of the $B$ mode.

This work was supported in part by the TMR Network ``Gravitational
Lensing: New Constraints on Cosmology and the Distribution of Dark
Matter'' of the EC under contract No. ERBFMRX-CT97-0172, the Canada
Foundation for Innovation funded computing infrastructure, and NSERC
grant 72013704.  We thank Francis Bernardeau, Peter Schneider for very
interesting and long discussions on skewness and other 3-pt estimators. 
We
thank Uros Seljak for the dark matter projection code, and Mike Jarvis
for correction equation (\ref{eqn:wabcd}).  TJZ would like
to thank CITA for its hospitality during his visit.

\bibliography{penbib}

\begin{thebibliography}{28}
\expandafter\ifx\csname natexlab\endcsname\relax\def\natexlab#1{#1}\fi

\bibitem[{{Bacon} {et~al.}(2002){Bacon}, {Massey}, {Refregier}, \&
  {Ellis}}]{2002astro.ph..3134B}
{Bacon}, D., {Massey}, R., {Refregier}, A., \& {Ellis}, R. 2002, in eprint
  arXiv:astro-ph/0203134, 3134--+

\bibitem[{{Bernardeau}(1998)}]{1998A&A...338..375B}
{Bernardeau}, F. 1998, \aap, 338, 375

\bibitem[{{Bernardeau} {et~al.}(2002){Bernardeau}, {Mellier}, \& {van
  Waerbeke}}]{2002A&A...389L..28B}
{Bernardeau}, F., {Mellier}, Y., \& {van Waerbeke}, L. 2002, \aap, 389, L28

\bibitem[{{Bernardeau} {et~al.}(1997){Bernardeau}, {van Waerbeke}, \&
  {Mellier}}]{1997A&A...322....1B}
{Bernardeau}, F., {van Waerbeke}, L., \& {Mellier}, Y. 1997, \aap, 322, 1

\bibitem[{{Bernardeau} {et~al.}(2003){Bernardeau}, {van Waerbeke}, \&
  {Mellier}}]{2003A&A...397..405B}
---. 2003, \aap, 397, 405

\bibitem[{{Brown} {et~al.}(2002){Brown}, {Taylor}, {Bacon}, {Gray}, {Dye},
  {Meisenheimer}, \& {Wolf}}]{2002astro.ph.10213B}
{Brown}, M.~L., {Taylor}, A.~N., {Bacon}, D.~J., {Gray}, M.~E., {Dye}, S.,
  {Meisenheimer}, K., \& {Wolf}, C. 2002, in eprint arXiv:astro-ph/0210213,
  10213--+

\bibitem[{{Crittenden} {et~al.}(2002){Crittenden}, {Natarajan}, {Pen}, \&
  {Theuns}}]{2002ApJ...568...20C}
{Crittenden}, R.~G., {Natarajan}, P., {Pen}, U., \& {Theuns}, T. 2002, \apj,
  568, 20

\bibitem[{{Davis} \& {Peebles}(1983)}]{1983ApJ...267..465D}
{Davis}, M. \& {Peebles}, P.~J.~E. 1983, \apj, 267, 465

\bibitem[{{Hamana} {et~al.}(2002{\natexlab{a}}){Hamana}, {Colombi}, {Thion},
  {Devriendt}, {Mellier}, \& {Bernardeau}}]{2002MNRAS.330..365H}
{Hamana}, T., {Colombi}, S.~T., {Thion}, A., {Devriendt}, J.~E.~G.~T.,
  {Mellier}, Y., \& {Bernardeau}, F. 2002{\natexlab{a}}, \mnras, 330, 365

\bibitem[{{Hamana} {et~al.}(2002{\natexlab{b}}){Hamana}, {Miyazaki},
  {Shimasaku}, {Furusawa}, {Doi}, {Hamabe}, {Imi}, {Kimura}, {Komiyama},
  {Nakata}, {Okada}, {Okamura}, {Ouchi}, {Sekiguchi}, {Yagi}, \&
  {Yasuda}}]{2002astro.ph.10450H}
{Hamana}, T., {Miyazaki}, S., {Shimasaku}, K., {Furusawa}, H., {Doi}, M.,
  {Hamabe}, M., {Imi}, K., {Kimura}, M., {Komiyama}, Y., {Nakata}, F., {Okada},
  N., {Okamura}, S., {Ouchi}, M., {Sekiguchi}, M., {Yagi}, M., \& {Yasuda}, N.
  2002{\natexlab{b}}, in eprint arXiv:astro-ph/0210450, 10450--+

\bibitem[{{Hockney} \& {Eastwood}(1988)}]{1988csup.book.....H}
{Hockney}, R.~W. \& {Eastwood}, J.~W. 1988, {Computer simulation using
  particles} (Bristol: Hilger, 1988)

\bibitem[{{Hoekstra} {et~al.}(2002){Hoekstra}, {Yee}, {Gladders}, {Barrientos},
  {Hall}, \& {Infante}}]{2002ApJ...572...55H}
{Hoekstra}, H., {Yee}, H.~K.~C., {Gladders}, M.~D., {Barrientos}, L.~F.,
  {Hall}, P.~B., \& {Infante}, L. 2002, \apj, 572, 55

\bibitem[{{Jain} {et~al.}(2000){Jain}, {Seljak}, \&
  {White}}]{2000ApJ...530..547J}
{Jain}, B., {Seljak}, U., \& {White}, S. 2000, \apj, 530, 547

\bibitem[{{Jarvis} {et~al.}(2002){Jarvis}, {Bernstein}, {Jain}, {Fischer},
  {Smith}, {Tyson}, \& {Wittman}}]{2002astro.ph.10604J}
{Jarvis}, M., {Bernstein}, G., {Jain}, B., {Fischer}, P., {Smith}, D., {Tyson},
  J.~A., \& {Wittman}, D. 2002, in eprint arXiv:astro-ph/0210604, 10604--+

\bibitem[{{Kaiser}(1987)}]{1987MNRAS.227....1K}
{Kaiser}, N. 1987, \mnras, 227, 1

\bibitem[{{Padmanabhan} {et~al.}(2002){Padmanabhan}, {Seljak}, \&
  {Pen}}]{2002astro.ph.10478P}
{Padmanabhan}, N., {Seljak}, U., \& {Pen}, U.~L. 2002, in eprint
  arXiv:astro-ph/0210478, 10478--+

\bibitem[{{Peacock} \& {Dodds}(1996)}]{1996MNRAS.280L..19P}
{Peacock}, J.~A. \& {Dodds}, S.~J. 1996, \mnras, 280, L19

\bibitem[{{Pen}(1998)}]{1998ApJ...498...60P}
{Pen}, U. 1998, \apj, 498, 60

\bibitem[{{Pen} {et~al.}(2002){Pen}, {Van Waerbeke}, \&
  {Mellier}}]{2002ApJ...567...31P}
{Pen}, U., {Van Waerbeke}, L., \& {Mellier}, Y. 2002, \apj, 567, 31

\bibitem[{{Refregier} {et~al.}(2002){Refregier}, {Rhodes}, \&
  {Groth}}]{2002ApJ...572L.131R}
{Refregier}, A., {Rhodes}, J., \& {Groth}, E.~J. 2002, \apjl, 572, L131

\bibitem[{{Schneider} \& {Lombardi}(2002)}]{2002astro.ph..7454S}
{Schneider}, P. \& {Lombardi}, M. 2002, in eprint arXiv:astro-ph/0207454,
  7454--+

\bibitem[{{Seljak} \& {Zaldarriaga}(1996)}]{1996ApJ...469..437S}
{Seljak}, U. \& {Zaldarriaga}, M. 1996, \apj, 469, 437+

\bibitem[{{Takada} \& {Jain}(2002)}]{2002astro.ph.10261T}
{Takada}, M. \& {Jain}, B. 2002, in eprint arXiv:astro-ph/0210261, 10261--+

\bibitem[{{Van Waerbeke} {et~al.}(2000){Van Waerbeke}, {Mellier}, {Erben},
  {Cuillandre}, {Bernardeau}, {Maoli}, {Bertin}, {Mc Cracken}, {Le F{\`e}vre},
  {Fort}, {Dantel-Fort}, {Jain}, \& {Schneider}}]{2000A&A...358...30V}
{Van Waerbeke}, L., {Mellier}, Y., {Erben}, T., {Cuillandre}, J.~C.,
  {Bernardeau}, F., {Maoli}, R., {Bertin}, E., {Mc Cracken}, H.~J., {Le
  F{\`e}vre}, O., {Fort}, B., {Dantel-Fort}, M., {Jain}, B., \& {Schneider}, P.
  2000, \aap, 358, 30

\bibitem[{{Van Waerbeke} {et~al.}(2002){Van Waerbeke}, {Mellier}, {Pell{\' o}},
  {Pen}, {McCracken}, \& {Jain}}]{2002A&A...393..369V}
{Van Waerbeke}, L., {Mellier}, Y., {Pell{\' o}}, R., {Pen}, U.-L., {McCracken},
  H.~J., \& {Jain}, B. 2002, \aap, 393, 369

\bibitem[{{Van Waerbeke} {et~al.}(2001){Van Waerbeke}, {Mellier}, {Radovich},
  {Bertin}, {Dantel-Fort}, {McCracken}, {Le F{\` e}vre}, {Foucaud},
  {Cuillandre}, {Erben}, {Jain}, {Schneider}, {Bernardeau}, \&
  {Fort}}]{2001A&A...374..757V}
{Van Waerbeke}, L., {Mellier}, Y., {Radovich}, M., {Bertin}, E., {Dantel-Fort},
  M., {McCracken}, H.~J., {Le F{\` e}vre}, O., {Foucaud}, S., {Cuillandre},
  J.-C., {Erben}, T., {Jain}, B., {Schneider}, P., {Bernardeau}, F., \& {Fort},
  B. 2001, \aap, 374, 757

\bibitem[{{White} \& {Hu}(2000)}]{2000ApJ...537....1W}
{White}, M. \& {Hu}, W. 2000, \apj, 537, 1

\bibitem[{{Zaldarriaga} \& {Scoccimarro}(2002)}]{2002astro.ph..8075Z}
{Zaldarriaga}, M. \& {Scoccimarro}, R. 2002, in eprint arXiv:astro-ph/0208075,
  8075--+

\end{thebibliography}
\bibliographystyle{apj}

\appendix

\end{document}